\documentclass[epj, nopacs]{svjour}

\usepackage{lineno}
\usepackage{graphicx}
\usepackage{amsmath}
\usepackage{amssymb}

\title{Dark-Photon Search using Data from CRESST-II Phase 2}

\author{
G.~Angloher\inst{1} \and
P.~Bauer\inst{1} \and
A.~Bento\inst{1,8} \and
C.~Bucci\inst{2} \and
L.~Canonica\inst{2,9} \and
X.~Defay\inst{3} \and
A.~Erb\inst{3,10} \and
F.~v.~Feilitzsch\inst{3} \and
N.~Ferreiro~Iachellini\inst{1} \and
P.~Gorla\inst{2} \and
A.~G\"utlein\inst{4,5,}\thanks{corresponding author, achim.guetlein@oeaw.ac.at} \and
D.~Hauff\inst{1} \and
J.~Jochum\inst{6} \and
M.~Kiefer\inst{1} \and
H.~Kluck\inst{4,5} \and
H.~Kraus\inst{7} \and
J.-C.~Lanfranchi\inst{3}
J.~Loebell\inst{6} \and
M.~Mancuso\inst{1} \and
A.~M\"unster\inst{3} \and
C.~Pagliarone\inst{2} \and
F.~Petricca\inst{1} \and
W.~Potzel\inst{3} \and
F.~Pr\"obst\inst{1} \and
R.~Puig\inst{4,5} \and
F.~Reindl\inst{1,}\thanks{Current address: INFN - Sezione di Roma, I-00185 Roma, Italy} \and
K.~Sch\"affner\inst{2,11} \and
J.~Schieck\inst{4,5} \and
S.~Sch\"onert\inst{3} \and
W.~Seidel\inst{1} \and
L.~Stodolsky\inst{1} \and
C.~Strandhagen\inst{6} \and
R.~Strauss\inst{1} \and
A.~Tanzke\inst{1} \and
H.H.~Trinh~Thi\inst{3} \and
C.~T\"urko\v{g}lu\inst{4,5} \and
M.~Uffinger\inst{6} \and
A.~Ulrich\inst{3} \and
I.~Usherov\inst{6} \and
S.~Wawoczny\inst{3} \and
M.~Willers\inst{3} \and
M.~W\"ustrich\inst{1} \and
A.~Z\"oller\inst{3}
}

\institute{
Max-Planck-Institut f\"ur Physik, D-80805 M\"unchen, Germany \and
INFN, Laboratori Nazionali del Gran Sasso, I-67010 Assergi, Italy \and
Physik-Department and Excellence Cluster Universe, Technische Universit\"at M\"unchen, D-85747 Garching, Germany \and
Institut f\"ur Hochenergiephysik der \"Osterreichischen Akademie der Wissenschaften, A-1050 Wien, Austria \and
Atominstitut, Vienna University of Technology, A-1020 Wien, Austria \and
Eberhard-Karls-Universit\"at T\"ubingen, D-72076 T\"ubingen, Germany \and
Department of Physics, University of Oxford, Oxford OX1 3RH, United Kingdom \\ \and
Also at: Departamento de Fisica, Universidade de Coimbra, P3004 516 Coimbra, Portugal \and
Also at: Massachusetts Institute of Technology, Cambridge, MA 02139, USA \and
Also at: Walther-Mei\ss{}ner-Institut f\"ur Tieftemperaturforschung, D-85748 Garching, Germany \and
Also at: GSSI-Gran Sasso Science Institute, 67100, L'Aquila, Italy
}

\newcommand{\FIG}[1]{Fig.~\ref{#1}}

\begin{document}

\abstract{
Identifying the nature and origin of dark matter is one of the major challenges for modern astro and particle physics. Direct dark-matter searches aim at an observation of dark-matter particles interacting within detectors. The focus of several such searches is on interactions with nuclei as provided e.g. by Weakly Interacting Massive Particles. However, there is a variety of dark-matter candidates favoring interactions with electrons rather than with nuclei. One example are dark photons, i.e., long-lived vector particles with a kinetic mixing to standard-model photons. In this work we present constraints on this kinetic mixing based on data from CRESST-II Phase 2 corresponding to an exposure before cuts of 52\,kg-days. These constraints improve the existing ones for dark-photon masses between 0.3 and 0.7\,keV/c$^2$.
}

\maketitle

\section{Introduction}
The dynamics of galaxies and galaxy clusters give strong hints for the existence of dark matter \cite{DM1}-\cite{DM3}. Recent measurements of the temperature fluctuations of the cosmic microwave background are well described with a dark-matter contribution of 26.6\,\% \cite{Planck} to the overall energy density of the universe. However, the nature and origin of dark matter is still unkown. Solving this dark-matter puzzle is one of the major challenges of modern astro and particle physics. 

Direct dark-matter searches \cite{CDEX}-\cite{XENON} aim at the observation of dark-matter particles interacting within detectors. Many of these experiments focus on interactions between dark-matter particles and nuclei as provided for example by Weakly Interacting Massive Particles (WIMPs) \cite{DM1}, \cite{DM2}. However, there is a variety of dark-matter models predicting particles which would favor interactions with electrons. One of these dark-matter candidates are dark photons \cite{DP1}-\cite{DAMIC_DP}, i.e., long-lived vector particles. The mass of these vector particles has to be smaller than two times the electron mass, otherwise they could decay into electron-positron pairs and their life time would be too small to be a dark-matter candidate.

The absorption of dark photons is similar to the photoelectric effect and the cross section $\sigma_V$ is approximately given by \cite{DP1}-\cite{DAMIC_DP}:
\begin{equation}\label{equ:CrossSection}
	\sigma_V(E_V = m_V c^2) v = \kappa^2 \sigma_{\gamma}(\hbar\omega = m_Vc^2) c,
\end{equation}
where $m_V$ is the mass of the dark photon, $v$ is their velocity,  $\kappa$ is
the kinetic mixing of dark photons and standard-model photons, and $\sigma_{\gamma}$ is the photoelectric cross-section for CaWO$_4$ depicted in \begin{figure}[htb]
	\centering
	\includegraphics[width=0.48\textwidth]{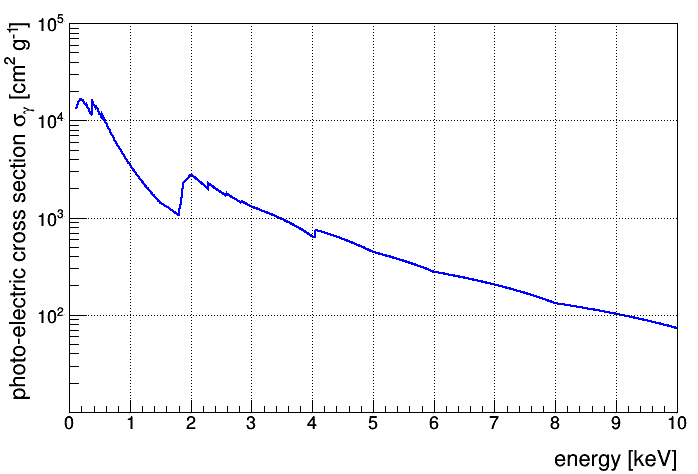}
	\caption{Photo-electric cross section $\sigma_{\gamma}$ for CaWO$_4$. The cross section for energies below 1\,keV was taken from \cite{CrossSection1}, for energies above 1\,keV from \cite{CrossSection2}.}
	\label{fig:CrossSection}
\end{figure}
\FIG{fig:CrossSection}.

When dark photons are absorbed inside the detectors of direct dark-matter searches the dark-photon energy is transferred to an electron similarly as for the photoelectric effect. Due to the short range of electrons in matter all energy is deposited within the detectors. Thus, the expected signal for the absorption of dark photons is a peak at the energy corresponding to the rest energy $m_V c^2$ of dark photons\footnote{The kinetic energy is negligible due to the small velocities of cold dark-matter.}. Assuming that dark matter consists only of dark photons, the absorption rate $R_S$ is given by \cite{DP1}-\cite{DAMIC_DP}:
\begin{equation}\label{equ:Rate}
	R_S = \frac{N_A}{A}\frac{\rho_{DM}}{m_V c^2} \cdot \kappa^2 \sigma_{\gamma}(\hbar \omega = m_V c^2) c,
\end{equation}
where $N_A$ is Avogadro's number, $A$ is the mass number of the target atoms, $\rho_{DM} = 0.3$\,GeV/cm$^3$ \cite{LocalDensity} is the local energy-density of dark matter.

Several direct dark-matter searches provide efficient methods for the discrimination between interactions with electrons and nuclei on an event-by-event basis. Since most backgrounds from natural radioactivity and cosmogenics interact with electrons, those methods allow a suppression of backgrounds for the search for dark-matter particles interacting with nuclei (e.g. WIMPs). However, since dark-photons also interact with electrons such background-sup\-pression methods cannot be applied.

Some dark-matter searches aim at an observation of an annual modulation of their event rate as it is for example expected for WIMPs \cite{CoGeNT}, \cite{DAMA}, \cite{XENON_modulation}. Since the absorption rate (equation (\ref{equ:Rate})) is independent of the dark-photon velocity, also this method cannot be applied to dark-photon searches. 

Nevertheless, direct dark-matter searches set the most stringent constraints for the kinetic mixing for dark-photon masses below $\sim 10$\,keV/c$^2$ \cite{DP1}-\cite{DAMIC_DP}.

\section{Data from CRESST-II Phase 2}\label{DataSelection}

CRESST-II is a direct dark-matter search using scintillating calcium tungstate (CaWO$_4$) crystals as detector material \cite{LiseResults}. In Phase 2 of CRESST-II (July 2013 - August 2015) 18 detector modules with a total detector mass of $\sim 5$\,kg were operated. For this work we only take into account data from the module with the lowest energy threshold of $0.307$\,keV. The same data set was also used to obtain the strongest limit on the cross section for spin-independent elastic scattering for masses of dark-matter particles $\lesssim 2$\,GeV/c$^2$ \cite{LiseResults}.

The detector module which obtained the data used for this work consists of two cryogenic detectors operated at $\mathcal{O}(10$\,mK). The so-called phonon detector based on a scintillating CaWO$_4$ crystal with a mass of $306$\,g measures the phonons generated by particle interactions within the crystal. The additional light detector is based on a silicon-on-sapphire (SOS) disc and measures the scintillation light emitted by the CaWO$_4$ crystal. Both detectors are surrounded by a scintillating and reflective housing to increase the light-collection efficiency. In addition, the scintillation light generated within the housing can be used for background suppression.

\begin{figure}[htb]
	\centering
	\includegraphics[width=0.48\textwidth]{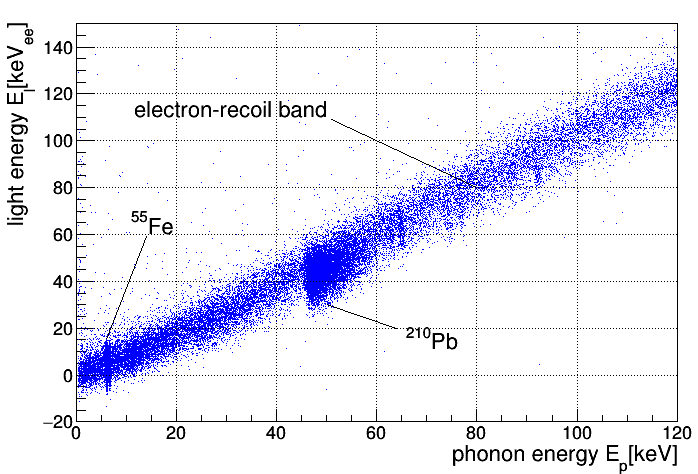}
	\caption{The light energy $E_l$, i.e., the scintillation light detected by the light detector, plotted against the phonon energy $E_p$. The electron-recoil band, where dark-photon events and most of the background events are expected, is clearly visible. The most prominent features within this band originate from an accidental irradiation with a ${}^{55}$Fe source (5.9 and 6.5\,keV) and in intrinsic contamination with ${}^{210}$Pb ($46.5$\,keV). Above the electron-recoil band so-called excess-light events are visible (see main text for further details).}
	\label{fig:LightEnergy_Energy}
\end{figure}
\FIG{fig:LightEnergy_Energy} depicts the data analyzed for this work after all cuts. The exposure before cuts is $52$\,kg-days. The events above the clearly visible electron-recoil band are called excess-light events. The origin of these events is not fully understood. However, a likely explanation are $\beta$ particles penetrating the detector housing before hitting the CaWO$_4$ crystal. Since the detector housing is also scintillating, such events would generate additional scintillation light resulting in a higher detected light-energy than expected for $\beta$ particles \cite{ExcessLight}. Only a fraction of 0.9\,\% of the total number of events are excess-light events. However, in the region of interest below 5\,keV this fraction increases to 9.2\,\%. Thus, excess-light events have to be taken into account as background component for this work.

\section{Search for dark-photon signals}

The expected signal from dark photons is a Gaussian peak within the electron-recoil band at the energy corresponding to the rest energy of dark photons. In this work we focus on dark-photon masses below 2\,keV/c$^2$. For larger masses other direct dark-matter searches have better sensitivities due to larger exposures and smaller background rates \cite{DP3}.

In order to save computation time and also to keep the empirical background model rather simple, we decided to take only events with phonon energies below 5\,keV into account for the Bayesian fits described in the next section. Due to the good energy resolution of the phonon detector of $0.062$\,keV (at 0.3\,keV) \cite{LiseResults} the influence of events with higher phonon energies on the resulting limits is negligible.

\subsection{Bayesian fits}
To study potential signals from dark photons we perform Bayesian fits of the potential signal and an empirical background model to the measured data. We decided on this fit method since systematic uncertainties and results of other measurements or sideband analyses can naturally be included into Bayesian methods. In addition, any problems with coverage as present for some Frequentist methods are avoided. The Bayes' theorem states that the posterior probability density function (PDF) is proportional to the product of the likelihood and the prior probability density functions of all parameters $\vec{\theta}$ of model $M$ (see e.g. \cite{Bayesian1}, \cite{Bayesian2}):
\begin{equation}\label{equ:Bayes}
P(M, \vec{\theta} | data ) = \frac{\mathcal{L}(data | M, \vec{\theta})P_0(\vec{\theta})}{\int d\vec{\theta}\mathcal{L}(data | M, \vec{\theta})P_0(\vec{\theta})}
\end{equation}
Bayesian fits maximize the posterior PDF with respect to the parameters $\vec{\theta}$ and provide a decent handling of systematic uncertainties and nuisance parameters. For a Bayesian fit we have to provide the likelihood function $\mathcal{L}(data | M, \vec{\theta})$ as well as prior PDFs for all parameters of our model.

\subsubsection{Likelihood}

In this work we use the following expression for the extended unbinned likelihood:
\begin{eqnarray}
	\mathcal{L}(data | M, \vec{\theta}) & = & \frac{\lambda^N}{N!} e^{-\lambda} \prod_{j=1}^{N} p(d_j | M, \vec{\theta}) \\
	p(d_j | M, \vec{\theta}) & = & \sum_i \frac{R_i}{\lambda} p_i(d_j | M, \vec{\theta}) \\
	\lambda & = & \sum_i R_i
\end{eqnarray}
where $N$ is the number of events in the fit range and $\lambda$ is the total expected rate which is given by the sum over the rates $R_i$ of the different signal and background components. The probability $p(d_j | M, \vec{\theta})$ that a data point $d_j$ is compatible with model $M$ and its parameters $\vec{\theta}$ is given by the sum over the probabilities $p_i(d_j | M, \vec{\theta})$ that $d_j$ belongs to component~$i$.

For all data points $d_j$ we take into account their phonon energy $E_{p,j}$ and their light energy $E_{l,j}$. The ratio between light and phonon energy is important to distinguish between excess-light events and normal electron-recoil events. This is of special importance for energies $\lesssim1$\,keV where the larger number of excess-light events would otherwise decrease the sensitivity of our dark-photon search.

The deposited energy of an event is split between phonon and light energy, where the light energy carries only a few percent of the total deposited energy. The energy calibration is done such that for a total deposited energy of 122\,keV caused by $\gamma$s from a ${}^{57}$Co calibration source the phonon and light energies are set to $E_p = 122$\,keV and $E_l = 122$\,keV$_{ee}$, respectively. With this calibration the phonon energy is equal to the deposited energy if the ratio between light and phonon energy is close to one. This is the case for events within the electron-recoil band. However, for nuclear-recoil events less light is generated and the phonon energy carries a larger fraction of the total deposited energy. This leads to a slight overestimation of the deposited energies. In \cite{TUM40} a correction for this effect is introduced. Since we are mainly dealing with electron-recoil events, we decided to not apply the correction from \cite{TUM40} for this work and assume that the total deposited energy is equal to the phonon energy. This approach is further supported by \cite{LiseResults} where no influence of the correction from \cite{TUM40} on the results could be found for the data set we are using for this work. 

With the assumption that the total deposited energy is given by the phonon energy, the probabilities $p_i(d_j | M, \vec{\theta}) = p_i(E_{p,j}, E_{l,j} | M, \vec{\theta})$ can be factorized\footnote{For better readability we omitted the indices $j$.}:
\begin{equation}
p_i(E_p, E_l | M, \vec{\theta}) = p_{p,i}(E_p|M,\vec{\theta})p_{l,i}(E_l|E_p,M,\vec{\theta}),
\end{equation}
where $p_{p,i}(E_p|M,\vec{\theta})$ are PDFs describing the distribution of phonon energies and $p_{l,i}(E_l | E_p, M,\vec{\theta})$ are PDFs describing the distribution of light energies for a given phonon energy. 

We will first discuss the PDFs $p_{p,i}(E_p|M,\vec{\theta})$ we used to describe the phonon energies of signal and background components. Afterwards we will describe in detail the corresponding PDFs $p_{l,i}(E_l|E_p,M,\vec{\theta})$ for the light energies.

For the expected dark-photon signal we use a Gaussian peak to model the distribution of the phonon energies. The position and the width of the Gaussian are fixed for each fit. However, we performed several fits where we varied the positions in 0.05\,keV steps from 0.3 to 2\,keV. For the width we use the energy dependent energy resolution of the phonon detector given by a linear interpolation between a resolution of 0.062\,keV at the threshold of 0.307\,keV and a resolution of 0.100\,keV at the 5.9\,keV line.

The empirical background model has three components. One component which describes the electron-recoil background from natural radioactivity and cosmogenics is modeled as a constant. The distribution of excess-light events is modeled by two components: One exponentially decaying function for events with small phonon energies $\lesssim2$\,keV and a constant for excess-light events with larger phonon energies.

We studied several different empirical parametrizations for the background model. The model presented here delivers the best description of the measured data in terms of Bayes factors, i.e., the ratios of the denominators of equation (\ref{equ:Bayes}). However, the limits for the kinetic mixing of dark photons are similar for all studied models.

All the PDFs $p_{p,i}(E_p|M,\vec{\theta})$ are multiplied with the energy dependent signal-survival probability \cite{LiseResults}, i.e., the probability that an event survives all cuts.
\begin{figure}
	\centering
	\includegraphics[width=0.48\textwidth]{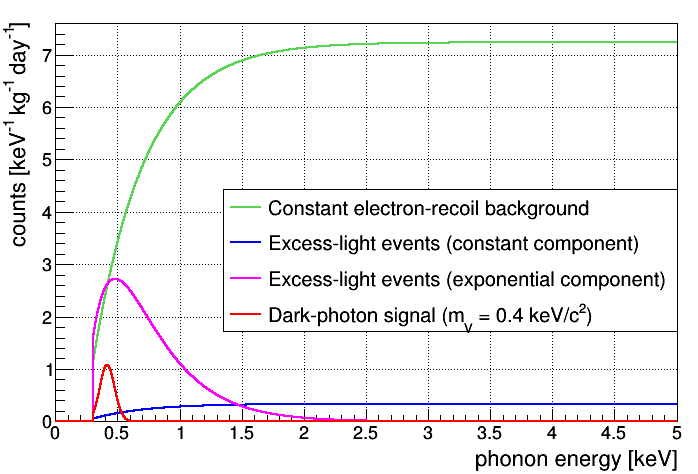}
	\caption{The PDFs $p_{p,i}(E_p|M,\vec{\theta})$ of all model components scaled by the respective rates obtained by the fit of the model to the data (compare section \ref{sec:FitResults})}
	\label{fig:FitComponents}
\end{figure}
\FIG{fig:FitComponents} shows the PDFs of all model components. For a meaningful comparison of the components we scaled all PDFs with the respective rates obtained by the fit (compare section \ref{sec:FitResults}). The influence of the energy dependent signal-survival probability is best visible for the constant background components.

To describe the PDFs for the light energy we use two different models, one for the electron-recoil band and one for excess-light events.

For events inside the electron-recoil band, i.e., the dark-photon signal and the constant background, the scintillation light is only generated within the CaWO$_4$ crystal. For a given phonon energy the statistical fluctuations of the light energies are caused by the baseline noise of the light detector and the counting statistics of the absorbed scintillation light. In principle, the light energies should follow a Poisson distribution. However, for large deposited energies a large number of photons is generated. Thus, the Poisson distribution can be approximated by a Gaussian. For small energies where this approximation would fail, the energy resolution of the light detector is dominated by the baseline noise. This noise is also very well approximated by a Gaussian (see also \cite{Para1}, \cite{Para2}). Thus, we use a Gaussian PDF for the description of the light energies of the electron-recoil band:
\begin{eqnarray}
	p_{\text{e-recoil}} (E_l | E_p, M, \vec{\theta}) & = & \frac{1}{\sigma(E_p)\sqrt{2\pi}} e^{-\frac{E_l-\mu(E_p)}{2\sigma(E_p)}} \label{equ:GammaBandPDF} \\
	\sigma(E_p) & = & \sqrt{S_0 + \mu(E_p)S_1} \\
	\mu(E_p) & = & P_0 \left(1 - P_1e^{-\frac{E_p}{P_2}}\right) \cdot E_p \label{equ:GammaBandMean}
\end{eqnarray}
where the mean $\mu(E_p)$ and the width $\sigma(E_p)$ depend on the phonon energy $E_p$. The parameters $S_0$ and $S_1$ are related to the baseline noise of the phonon and light detectors and the counting statistics of absorbed scintillation light, respectively. The parameter $P_0$ describes the proportionality of the scintillation light with respect to the phonon energy. Due to the calibration of the light energies, $P_0$ is close to 1. The parameters $P_1$ and $P_2$ describe the deviation from proportionality for low energies $E_p$. This so-called non-proportionality effect \cite{NP1}-\cite{NP3} is related to the energy-dependent stopping power of electrons, i.e., $dE/dx$ is larger for electrons with low energies leading to a decreased light output \cite{NP1}-\cite{NP3}. 

Excess-light events are not described by the already mentioned Gaussian PDF for the electron-recoil band. As already stated in section~\ref{DataSelection}, we assume that excess light events originate from external $\beta$s traversing the scintillating detector housing before hitting the CaWO$_4$ crystal. Within this model, the additional scintillation light generated within the housing is added to the normal scintillation light generated by $\beta$s hitting the CaWO$_4$ crystal. The statistical fluctuations of the latter follow the distribution given by equation (\ref{equ:GammaBandPDF}). We use an exponentially decaying function to model the additional scintillation light which is generated when particles penetrate the housing.\footnote{In principle, the distribution of this contribution should follow a Landau distribution. However, after the convolution with a Gaussian the Landau distribution and an exponential decay have similar shapes. Since an exponential leads to a simpler solution of the convolution integral, we decided to use an exponential distribution.} The distribution of the sum of two random variables follows the convolution of the distributions of the two random variables (see e.g. \cite{Statistics}). Thus, the distribution of the light energies for excess-light events is given by:
\begin{eqnarray*}
& & p_{\text{excess}}(E_l|E, M, \vec{\theta}) = p_{\text{additional}} \ast p_{\text{e-recoil}}\\
& = &\int_{0}^{\infty} \frac{1}{E_{l,0}} e^{-\frac{E_l'}{E_{l,0}}} \frac{1}{\sigma(E)\sqrt{2\pi}} e^{-\frac{(E_l-E_l'-\mu(E))^2}{2\sigma(E)}} dE_l' \\
\end{eqnarray*}
where $E_{l,0}$ is the slope of the exponential decay for the additional scintillation light, $\mu(E)$ and $\sigma(E)$ are the mean and width of the PDF describing the electron-recoil band.

\subsubsection{Prior distributions}

In Bayesian statistics the prior distributions of the model parameters are an important input. These prior distributions model the knowledge on each parameter before the experiment. Typically, the prior distributions are based on previous experiments, side-band analysis, or theoretical predictions.

For this work we have chosen uniform priors for most parameters\footnote{i.e., rates of signal and background components, decay constants for the exponentials in phonon and light energy for escess-light events, and $S_0$ and $S_1$ describing the width of the light-yield distribution of electron-recoil events.} to model our ignorance of these parameters. Uniform PDFs can only be defined for a finite range. Thus, we chose ranges which include the majority of the posterior PDFs for all parameters. Of course, we excluded unphysical regions of the parameter space (e.g. negative rates).

Only for the parameters $P_i$ of equation~(\ref{equ:GammaBandMean}) we use Gaussians instead of uniform PDFs. Parameter $P_2$ describes the energy scale of the non-proportionality effect. For our data set this parameter is $\sim 20$\,keV. Thus, the exponential decay of the mean of the electron-recoil band is not pronounced in our fit region below 5\,keV. In order to obtain meaningful prior distributions, we performed a dedicated fit of the events with energies above 5\,keV. As a result of this fit we use Gaussian PDFs centered around the best-fit values and widths of $\sim 10$\,\% for the prior distributions of the parameters $P_i$.

\subsection{Fit results}\label{sec:FitResults}

\begin{figure}[htb]
	\centering
	\includegraphics[width=0.48\textwidth]{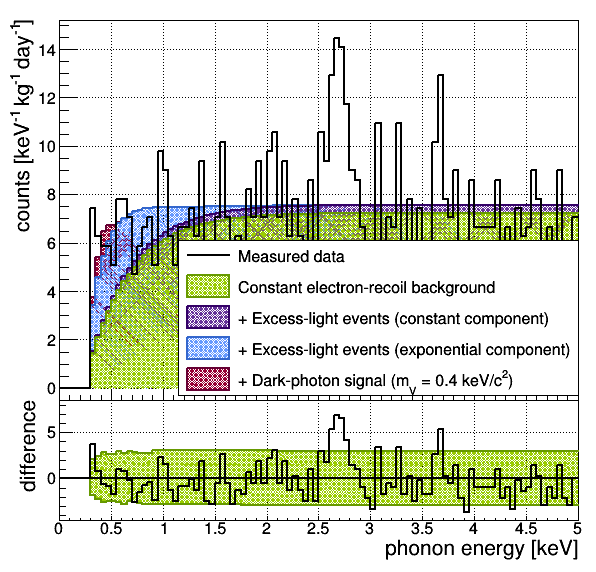}
	\caption{\textit{Top panel:} Fit results for the phonon energies. In addition to the measured data, the stacked contributions of all components are shown. The fixed mean of the signal peak was set to 0.4\,keV. The drop in the models below 2\,keV is related to the decreasing signal-survival probability. \textit{Bottom panel:} The differences between the data and the fitted model in each bin are depicted as solid black line. In addition, the statistical uncertainties (central 90\,\% region) of the fitted model are shown as green-shaded region.}
	\label{fig:FitEnergy}
\end{figure}

\begin{figure}[htb]
	\centering
	\includegraphics[width=0.48\textwidth]{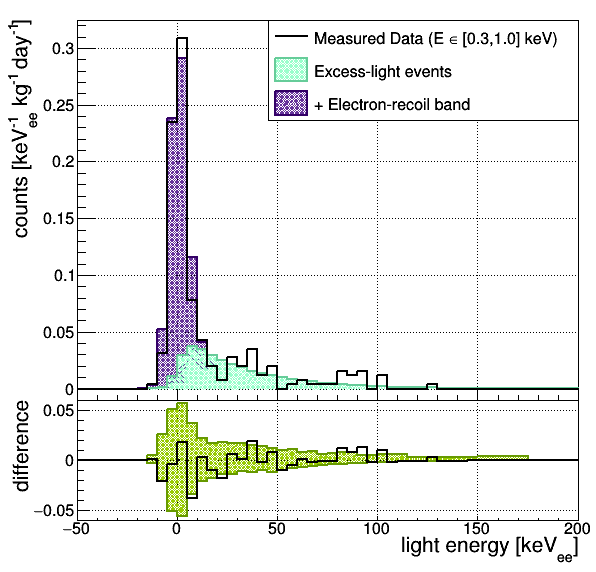}
	\caption{\textit{Top panel:} Fit results for the light energies. Only events within the phonon-energy slice between 0.3 and 1\,keV are shown in this plot. In addition to the measured data, the stacked contributions from the electron-recoil band and excess-light events are shown. \textit{Bottom panel:} The differences between the data and the fitted model are depicted as solid black line. In addition, the statistical uncertainties (central 90\,\% region) of the fitted model are shown as green-shaded region.}
	\label{fig:FitLightEnergy}
\end{figure}

\FIG{fig:FitEnergy} and \FIG{fig:FitLightEnergy} show the results of a Bayesian fit for a fixed position of 0.4\,keV for the signal peak. \FIG{fig:FitEnergy} shows the phonon energies of all events in the region used for fitting. \FIG{fig:FitLightEnergy} shows the light energies of the events within the phonon-energy slice between 0.3\,keV and 1\,keV. In both panels a good agreement between the data and the fit model becomes evident. The origin of the peak between 2.5 and 3\,keV is not fully understood and, thus, we conservatively did not include this peak into the background model. This leads to the discrepancies between data and model visible in \FIG{fig:FitEnergy}. However, it should be mentioned that this peak has no influence on the result of the peak search performed only for energies below 2\,keV.  

\begin{figure}[htb]
	\centering
	\includegraphics[width=0.49\textwidth]{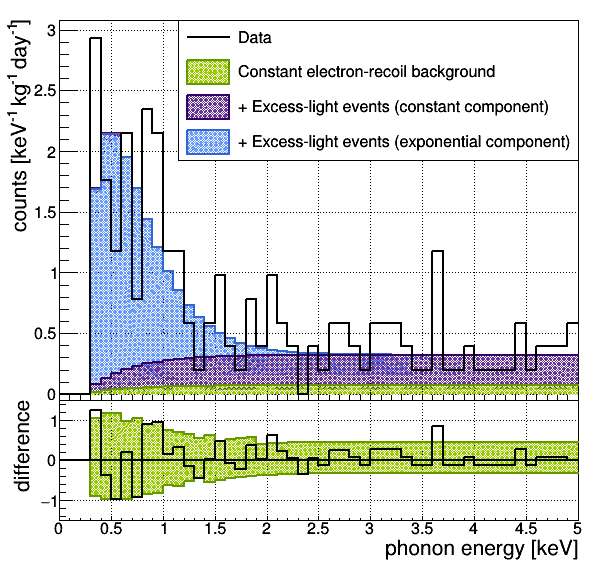}
	\caption{\textit{Top panel:} Distribution of phonon energies of excess-light events. For this plot only events with light energies larger than the 99\,\% quantile of the electron-recoil band (equation~(\ref{equ:GammaBandPDF})) were taken into account. In addition, the stacked distributions of all components of the fit model are shown.  The contribution by the signal is too small to be visible in this plot. The drop in the models below 2\,keV is related to the decreasing signal-survival probability. \textit{Bottom panel:} The differences between the data and the fitted model are depicted as solid black line. In addition, the statistical uncertainties (central 90\,\% region) of the fitted model are shown as green-shaded region.}
	\label{fig:ExcessLightSpectrum}
\end{figure}
\FIG{fig:ExcessLightSpectrum} depicts the distribution of the phonon energies for excess-light events. For this plot only events with light energies above the 99\,\% quantile of the electron-recoil band (equation~(\ref{equ:GammaBandPDF})) were taken into account. It is clearly visible that the excess-light events are well described by the fitted model.

\section{New limit for kinetic mixing}

Before we can set a limit on the kinetic mixing of dark photons we first have to estimate a limit on the signal rate $R_S$, i.e., the rate of the signal component of the fit model. Therefore, we first have to create the marginalized posterior PDF of the signal rate by integrating the posterior PDF over all other parameters:
\begin{equation}
	P(R_S|data) = \int P(M, \vec{\theta} | data ) d\vec{\theta} \biggr\rvert_{\theta \neq R_{S}}
\end{equation}

\begin{figure}[htb]
	\centering
	\includegraphics[width=0.49\textwidth]{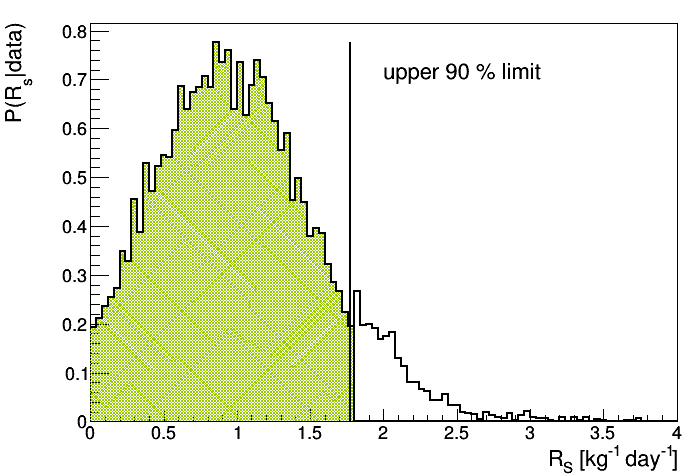}
	\caption{Marginalized posterior PDF for the signal rate $R_S$. For this example with a fixed peak position of 0.4\,keV the upper 90\,\% limit is $1.77$\,counts\,kg$^{-1}$day$^{-1}$.}
	\label{fig:Marginalized}
\end{figure}
\FIG{fig:Marginalized} shows the marginalized posterior PDF for the signal rate $R_S$ for a peak position of 0.4\,keV as an example. This marginalized PDF was obtained by a Markov Chain Monte Carlo algorithm \cite{Bayesian2} which was used to perform all fits for this work.

An upper limit for the signal rate can be obtained as the corresponding quantile of $P(R_S|data)$, e.g., the upper 90\,\% limit for $R_S$ corresponds to the 90\,\% quantile of $P(R_S|data)$. For the example of \FIG{fig:Marginalized} this upper 90\,\% limit is $~1.77$\,counts\,kg$^{-1}$day$^{-1}$.

This limit on the signal rate $R_S$ can be converted into a limit on the kinetic mixing using equation (\ref{equ:Rate}). We repeated the described Bayesian fit-procedure with different fixed positions for the dark-photon signal in 0.05\,keV steps from 0.3 to 2\,keV. The resulting limit for the kinetic mixing is shown in
\begin{figure}[htb]
	\centering
	\includegraphics[width=0.49\textwidth]{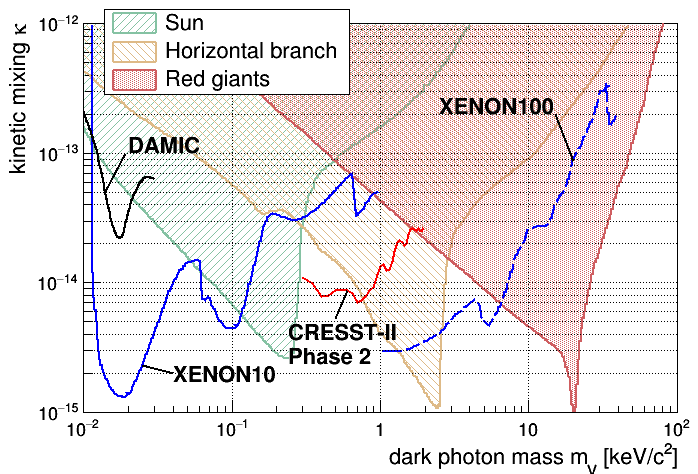}
	\caption{Upper limits on the kinetic mixing of dark photons. Limits from anomalous energy loss in the sun, horizontal-branch stars, and red giants are shown as shaded regions \cite{DP3}. Limits from XENON10 and XENON100 \cite{DP3} are shown as (blue) solid and dashed lines, respectively. The limit from DAMIC \cite{DAMIC_DP} is shown as solid black line. Our limit obtained from data from CRESST-II Phase 2 is depicted as a (red) solid line.}
	\label{fig:Limit}
\end{figure}
\FIG{fig:Limit} as a function of dark-photon mass. In addition, existing limits (90\,\% confidence level) from astronomy, the XENON and DAMIC experiments are shown. Our result improves the existing constraints for dark-photon masses between 0.3 and 0.7\,keV/c$^2$.

The recently started (July 2016) Phase 1 of CRESST-III has the potential to further improve this limit. The detectors operated in this phase of CRESST-III will have energy thresholds of $\lesssim0.1$\,keV \cite{CRESST_3}. Thus, with these detectors we can extend our limits towards smaller dark-photon masses of $\lesssim0.1$\,keV/c$^2$. 

\section{Conclusions}

The dynamics of galaxies and galaxy clusters give evidence for the existence of dark matter. However, its origin and nature remain unknown up to now. There is a variety of theories for dark matter. In recent years, theories predicting interactions of dark-matter particles with electrons rather than nuclei became more popular. One example are dark photons, i.e., long-lived vector particles with a kinetic mixing to standard-model photons.

Like several other direct dark-matter searches, CRESST-II is optimized for an observation of dark-matter particles interacting with nuclei. However, the obtained data can also be used to search for dark-matter candidates with different interactions. In this work we present the limits for the kinetic mixing of dark photons based on data from Phase 2 of CRESST-II corresponding to an exposure of $52$\,kg-days. To obtain this limit we performed Bayesian fits of an empirical background model and a potential dark-photon signal to the measured data. Our new limit improves the existing constraints for dark-photon masses between 0.3 and 0.7\,keV/$c^2$. Due to its low energy thresholds, the recently started CRESST-III Phase 1 has the potential to further improve this limits.

\section*{Acknowledgements }
We are grateful to LNGS for their generous support of CRESST, in particular to Marco Guetti for his
constant assistance. This work was supported by the DFG cluster of excellence: Origin and Structure
of the Universe, by the Helmholtz Alliance for Astroparticle Physics, and by the BMBF: Project
05A11WOC EURECA-XENON.

In addition, we would like to thank J. Pradler for helpful discussions about the physics of dark photons. Also we would like to thank M. T\"uchler and L. Lechner for their help with the software used to perform the Bayesian fits.

\end{document}